\documentclass[12pt]{article}
\usepackage{graphicx}

\font\grande=cmr9.5 scaled \magstep4
\font\medio=cmr9.5 scaled \magstep2
\outer\def\beginsection#1\par{\medbreak\bigskip
      \message{#1}\leftline{\bf#1}\nobreak\medskip
\vskip-\parskip
      \noindent}

\begin{document}
\bibliographystyle {unsrt}

\titlepage
\begin{flushright}
CERN-PH-TH/2009-062
\end{flushright}
\vspace{15mm}
\begin{center}
{\grande Ohmic currents and pre-decoupling magnetism}
\vskip1.1cm
 
Massimo Giovannini$^{a,b}$ and  Nguyen Quynh Lan$^{c}$
 
\vskip1.1cm
{\sl $^a$  Department of Physics, Theory Division, CERN, 1211 Geneva 23, Switzerland}
\vskip 0.2cm
{\sl $^b$ INFN, Section of Milan-Bicocca, 20126 Milan, Italy}
\vskip 0.2cm
{\sl $^c$  Hanoi National University of Education, 136 Xuan Thuy, Cau Giay, Hanoi, Vietnam}
\vspace{6mm}
\end{center}

\vskip 2cm
\centerline{\medio  Abstract}
Ohmic currents induced prior to decoupling are investigated in a standard 
transport model accounting both for the expansion of the background geometry as well as of its relativistic inhomogeneities.  
The relative balance of the Ohmic electric fields in comparison with the  Hall and thermoelectric contributions is specifically addressed. The impact of the Ohmic currents on the evolution of curvature perturbations is discussed numerically and it is shown to depend explicitly upon the evolution of the conductivity. 
\noindent

\vspace{15mm}
\vfill
\newpage
Prior to photon decoupling the plasma is electrically neutral, the static (Coulomb) potential is exponentially suppressed beyond the Debye length (see, e.g. \cite{boyd})
while  the concentration of the electric charges is $10^{-10}$ times smaller than the concentration of the photons (see e.g. \cite{max1}). This effect would naively seem to  increase the role of the Hall and thermoelectric terms whose 
magnitude is inversely proportional to the charge concentration \cite{spitzer}.  Ohmic electric fields might also be induced because of the presence of large-scale magnetic fields. The value of the conductivity is then crucial for determining the magnetic and electric diffusivity scales. The aim of the present paper is to clarify the situation and investigate more quantitatively  
the different contributions responsible of Ohmic currents\footnote{For cold plasmas in the laboratory the so-called Ohm law (see, e.g. \cite{boyd}) is a relation  (often non-linear)  involving 
the total current the electric and magnetic fields, the electron and ion pressures, the bulk velocity of the plasma. In curved backgrounds, on top of the previous quantities, one needs to take into account the effects due to the expansion and to the large-scale fluctuations of the geometry. 
The resulting Ohm law is then, effectively, an evolution equation for the Ohmic current. 
}  especially in the light of the ongoing attempt of a consistent inclusion of large-scale magnetic fields in the calculation of the Cosmic Microwave Background (CMB) observables \cite{max2,max3}. 
Consider, to begin with,
the Vlasov-Landau system of equations for electrons and 
ions\footnote{The conformal 
 time coordinate will be denoted by $\tau$ and  the geometry will be assumed to be conformally flat, i.e. $g_{\mu\nu} = a^2(\tau) \eta_{\mu\nu}$, where $\eta_{\mu\nu} = \mathrm{diag}(1,\,-1,\, -1,\, -1)$ 
 is the Minkowski metric.}
\begin{equation}
  \frac{\partial f_{\mathrm{e,i}}}{\partial \tau}  + \vec{v} \cdot \vec{\nabla}_{\vec{x}} f_{\mathrm{e,i}} \mp e [ \vec{E} + \vec{v} \times \vec{B}] \cdot \vec{\nabla}_{\vec{q}} f_{\mathrm{e,i}} =\biggl( \frac{\partial f_{\mathrm{e,i}}}{\partial \tau }\biggr)_{\mathrm{coll}},
\label{VL1}
\end{equation}
where $\vec{E}=a^2 \vec{{\mathcal E}}$ and $\vec{B}=a^2 \vec{{\mathcal B}}$ are, respectively, 
 the comoving electric and magnetic fields; $\vec{v} = \vec{q}/\sqrt{q^2 + m^2 a^2}$ is the velocity and $\vec{q}$ is the comoving three-momentum. In the ultra-relativistic limit (i.e. $q \gg m a$), $\vec{v} = \vec{q}/|\vec{q}|$ and, therefore,  Eq. (\ref{VL1}) is invariant under a Weyl rescaling of the geometry $g_{\mu\nu}$: this boils down to the conclusion that, absent the relativistic fluctuations of the geometry (which will be introduced in a moment) the Vlasov-Landau system has the same form it would have in flat space-time provided the underlying background geometry is spatially  flat. Conversely, $\vec{q} = m a \vec{v}$ when the given species are non-relativistic;  Weyl invariance is then broken by the masses of the electrons and of the ions (i.e., respectively,  $m_{\mathrm{e}}$ and $m_{\mathrm{i}} \simeq m_{\mathrm{p}}$).  
In terms of the distribution functions of Eq. (\ref{VL1}), the evolution equations of the electromagnetic fields are given by: 
\begin{eqnarray}
&&\vec{\nabla}\cdot \vec{E} = 4\pi e \int d^{3}v [f_{\mathrm{i}}(\vec{x}, \vec{v}, \tau) - f_{\mathrm{e}}(\vec{x}, \vec{v}, \tau)], \qquad \vec{\nabla}\cdot \vec{B} =0,
\label{MX1}\\
&& \vec{\nabla} \times \vec{E}  + \vec{B}' = 0, \qquad \vec{\nabla}\times \vec{B} - \vec{E}'= 
4\pi e \int d^{3}v\,\vec{v}\,[f_{\mathrm{i}}(\vec{x}, \vec{v}, \tau) - f_{\mathrm{e}}(\vec{x}, \vec{v}, \tau)], 
\label{MX2}
\end{eqnarray}
where the prime denotes a derivation with respect to 
the conformal time coordinate $\tau$.
The evolution equations of the comoving concentrations of electrons and ions (i.e. respectively $n_{\mathrm{e}}$ and $n_{\mathrm{i}}$)
can  be written, in explicit terms, as \footnote{Throughout the paper the physical quantities will be denoted by a tilde while the comoving quantities will appear 
without the tilde. For instance the (comoving) concentrations, energy densities 
and pressures of electrons and ions will be denoted, respectively, by $n_{\mathrm{e,i}} = a^3 \tilde{n}_{\mathrm{e,i}}$, $\rho_{\mathrm{e,i}} = a^3 \rho_{\mathrm{e,i}}$ and by 
$p_{\mathrm{e,\,i}} = n_{\mathrm{e,\,i}} T_{\mathrm{e,\,i}} = a^4 \tilde{p}_{\mathrm{e,\,i}}$. Similarly, for photons,  $\rho_{\gamma} = \
(\pi^2/15) T_{\gamma}^4 =a^4 \tilde{\rho}_{\gamma}$. When needed these two notations will be employed without further explanations.}
\begin{equation}
\frac{\partial n_{\mathrm{e}}}{\partial \tau} + \vec{\nabla} \cdot( n_{\mathrm{e}} 
\vec{v}_{\mathrm{e}}) - 3 n_{\mathrm{e}} \psi' =0, \qquad 
\frac{\partial n_{\mathrm{i}}}{\partial \tau} + \vec{\nabla} \cdot( n_{\mathrm{i}} 
\vec{v}_{\mathrm{i}}) - 3 n_{\mathrm{i}} \psi' =0,
\label{con1}
\end{equation}
where $\psi$ is the (scalar) fluctuation of spatial components of the metric in the longitudinal gauge \cite{bardeen} defined by the conditions $\delta_{\mathrm{s}} g_{00} = 2 a^2 \phi$, 
$\delta_{\mathrm{s}} g_{ij} = 2 a^2 \psi \, \delta_{ij}$. Introducing 
the global charge and the total current, i.e. 
\begin{equation}
\rho_{\mathrm{q}} = e (n_{\mathrm{i}} - n_{\mathrm{e}}),\qquad \vec{J} = e (n_{\mathrm{i}} \vec{v}_{\mathrm{i}} - n_{\mathrm{e}} \vec{v}_{\mathrm{e}}),  
\label{con2}
\end{equation}
the difference of the two equations reported in Eq. (\ref{con1})  implies that $\rho_{\mathrm{q}}' +  \vec{\nabla} \cdot \vec{J} -3 \psi' \rho_{\mathrm{q}}=0$. Using Eq. (\ref{con2}) the relevant  Maxwell equations become $\vec{\nabla}\cdot \vec{E} = 4 \pi \rho_{\mathrm{q}}$ and $\vec{\nabla}\times \vec{B} = 4\pi \vec{J} + \vec{E}\,'$.
Recalling that the pre-decoupling plasma is globally neutral, i.e.  $n_{\mathrm{e}} = n_{\mathrm{i}} = \eta_{\mathrm{b}} n_{\gamma}$ 
where $n_{\gamma}$ is the comoving concentration of photons and $\eta_{\mathrm{b}}$ is the ratio between 
the baryonic concentration and the photon concentration, i.e. $\eta_{\mathrm{b}} = 6.29 \times 10^{-10}\, (h_{0}^2 \Omega_{\mathrm{b}0}/0.02273) [T_{\gamma  0}/(2.725 ~\mathrm{K})]^{-3}$ where $\Omega_{\mathrm{b}0}$ is the critical fraction of baryons and $T_{\gamma 0}$ is the CMB temperature. The fiducial values of the cosmological parameters employed to illustrate the present estimates 
correspond to the best fit of the WMAP 5yr data alone \cite{WMAP5a,WMAP5b}. 
The conductivity (and the related mobility) can be computed in the customary framework of the Krook model \cite{krook1,krook2} which holds for weakly ionized plasmas, and, with some numerical differences, also in the fully ionized case. The collision terms of Eq. (\ref{VL1})  can then be written as 
\begin{equation}
 \biggl(\frac{\partial f_{\mathrm{e}}}{\partial \tau }\biggr)_{\mathrm{coll}} = \Gamma_{\mathrm{ei}} ( f_{\mathrm{e}} - \overline{f}_{\mathrm{e}}), \qquad \biggl(\frac{\partial f_{\mathrm{i}}}{\partial \tau}\biggr)_{\mathrm{coll}} = \Gamma_{\mathrm{ie}} ( f_{\mathrm{i}} - \overline{f}_{\mathrm{i}}),
\label{COLL1}
\end{equation}
where $\Gamma_{\mathrm{ei}}$ and $\Gamma_{\mathrm{ie}}$ are the collision rates of electrons and ions and where 
$\overline{f}_{\mathrm{e,i}}$ are two Maxwellian distributions, i.e. $\overline{f}(v)= [m a/(2 \pi T)]^{3/2} \exp{[- m a v^2 \, a/(2 T)]}$.
The induced electric field slightly perturb the Maxwellian distributions and, therefore, 
the explicit form of the conductivity can be 
derived from Eq. (\ref{MX2}) by following exactly the same steps of the standard calculation plasma calculation 
(see e.g. \cite{boyd,spitzer}) with the important difference that, because of  the breaking of Weyl invariance, the scale factors 
$a(\tau)$ appear ubiquitously:
\begin{eqnarray}
&& \sigma = \frac{9}{8 \pi \sqrt{3}} \frac{T}{e^2} \sqrt{\frac{T}{m_{\mathrm{e}} a}} [\ln{\Lambda_{\mathrm{C}}(T)}]^{-1} = 
4.35 \times 10^{-7}\, \mathrm{eV} \biggl(\frac{T_{\gamma 0}}{2.725\, \mathrm{K}}\biggr)^{3/2} \biggl(\frac{h_{0}^2 \Omega_{\mathrm{M}0}}{0.1326}\biggr)^{1/2} \sqrt{\frac{a_{\mathrm{eq}}}{a}},
\nonumber\\
&& \Lambda_{\mathrm{C}}(T) = \frac{3}{2 e^3} \biggl(\frac{T^3}{\pi n_{0}}\biggr)^{1/2} =  1.102\times 10^{8} \biggl(\frac{h_{0}^2 \Omega_{\mathrm{b}0}}{0.02273}\biggr)^{-1/2},
\label{COLL4}
\end{eqnarray}
where $n_{0}$ is the common value of the (comoving) electron and ion concentrations; $\Lambda_{\mathrm{C}}$ is the argument of the Coulomb logarithm and  $\Omega_{\mathrm{M}0}$ is the critical fraction of matter in the $\Lambda$CDM model. We are now interested in the evolution equation of the Ohmic current  $\vec{J}$ whose 
explicit form can be derived by combining the governing equations for electrons and ions: 
\begin{eqnarray}
&& \vec{v}_{\mathrm{i}}\,' + {\mathcal H} \vec{v}_{\mathrm{i}} = 
\frac{e n_{\mathrm{i}}}{\rho_{\mathrm{i}} a} [ \vec{E} + \vec{v}_{\mathrm{i}} \times 
\vec{B}] - \vec{\nabla} \phi - \frac{\vec{\nabla} p_{\mathrm{i}}}{a \rho_{\mathrm{i}}}
+ a \Gamma_{\mathrm{ie}}  \frac{\rho_{\mathrm{e}}}{\rho_{\mathrm{i}}} (\vec{v}_{\mathrm{e}} - \vec{v}_{\mathrm{i}}) + 
\frac{4}{3} \frac{\tilde{\rho}_{\gamma}}{ \tilde{\rho}_{\mathrm{i}}} a \Gamma_{\mathrm{i}\gamma} (\vec{v}_{\gamma} - \vec{v}_{\mathrm{i}}),
\label{COLL5}\\
&& \vec{v}_{\mathrm{e}}\,' + {\mathcal H} \vec{v}_{\mathrm{e}} = 
-\frac{e n_{\mathrm{e}}}{\rho_{\mathrm{e}} a} [ \vec{E} + \vec{v}_{\mathrm{e}} \times 
\vec{B}] - \vec{\nabla} \phi - \frac{\vec{\nabla} p_{\mathrm{e}}}{a \rho_{\mathrm{e}}}
+a \Gamma_{\mathrm{ei}} (\vec{v}_{\mathrm{i}} - \vec{v}_{\mathrm{e}}) +  
\frac{4}{3} \frac{\tilde{\rho}_{\gamma}}{ \tilde{\rho}_{\mathrm{e}}} 
a \Gamma_{\mathrm{e}\gamma} (\vec{v}_{\gamma} - \vec{v}_{\mathrm{e}}).
\label{COLL6}
\end{eqnarray}
By taking the difference of Eq. (\ref{COLL5}) (multiplied by $e\,n_{\mathrm{i}}$) and of Eq. (\ref{COLL6}) (multiplied by $e\,n_{\mathrm{e}}$) the following equation can be obtained:
\begin{eqnarray}
&& \frac{\partial \vec{J}}{\partial \tau} + {\mathcal H} \vec{J} = \frac{\omega_{\mathrm{pe}}^2 + \omega_{\mathrm{pi}}^2}{4\pi} \vec{E} - 
e (n_{\mathrm{i}} - n_{\mathrm{e}}) \vec{\nabla}\phi - 
e n_{\mathrm{i}} \frac{\vec{\nabla} p_{\mathrm{i}}}{a \rho_{\mathrm{i}}} + 
e n_{\mathrm{e}} \frac{\vec{\nabla} p_{\mathrm{e}}}{a \rho_{\mathrm{e}}}
\nonumber\\
&& + e n_{\mathrm{e}} \, a\,\Gamma_{\mathrm{ei}} \biggl(1 + \frac{m_{\mathrm{e}}}{
m_{\mathrm{i}}}\biggr)\biggr[ \frac{(n_{\mathrm{i}} - n_{\mathrm{e}}) (m_{\mathrm{e}} + m_{\mathrm{i}})}{m_{\mathrm{i}} n_{\mathrm{e}} + n_{\mathrm{i}} m_{\mathrm{e}}} \vec{v}_{\mathrm{b}} - \frac{(m_{\mathrm{i}} + m_{\mathrm{e}})}{e (n_{\mathrm{i}}
 m_{\mathrm{e}} + n_{\mathrm{e}} m_{\mathrm{i}})} \vec{J} \biggr]
 \nonumber\\
 && + \frac{e^2 n_{\mathrm{e}} n_{\mathrm{i}}(m_{\mathrm{e}} + m_{\mathrm{i}})}{ 
 a m_{\mathrm{e}}(n_{\mathrm{i}} m_{\mathrm{e}} + n_{\mathrm{e}} m_{\mathrm{i}})}\biggl(1 + \frac{m_{\mathrm{e}}}{m_{\mathrm{i}}}\biggr) \vec{v}_{\mathrm{b}} \times \vec{B} + \frac{e}{(n_{\mathrm{i}} m_{\mathrm{e}} + n_{\mathrm{e}} m_{\mathrm{i}})a}
\biggl(n_{\mathrm{i}} \frac{m_{\mathrm{e}}}{m_{\mathrm{i}}} - n_{\mathrm{e}} 
\frac{m_{\mathrm{i}}}{m_{\mathrm{e}}}\biggr) \vec{J}\times \vec{B}
\nonumber\\
&& + \frac{4}{3} e \rho_{\gamma} \biggl\{ \biggl( \frac{\Gamma_{\mathrm{i}\gamma}}{m_{\mathrm{i}}} - 
\frac{\Gamma_{\mathrm{e} \gamma}}{m_{\mathrm{e}}}\biggr) \vec{v}_{\gamma} 
+ \biggl[ \frac{\Gamma_{\mathrm{e}\gamma} n_{\mathrm{i}} (m_{\mathrm{e}} + m_{\mathrm{i}})}{m_{\mathrm{e}} ( 
m_{\mathrm{i}} n_{\mathrm{e}} + m_{\mathrm{e}} n_{\mathrm{i}})} - 
\frac{\Gamma_{\mathrm{i}\gamma} n_{\mathrm{e}} (m_{\mathrm{e}} + m_{\mathrm{i}})}{m_{\mathrm{i}} ( 
m_{\mathrm{i}} n_{\mathrm{e}} + m_{\mathrm{e}} n_{\mathrm{i}})} \biggr] \vec{v}_{\mathrm{b}} 
\nonumber\\
&& - \biggl[ \frac{\Gamma_{\mathrm{e}\gamma}}{e (m_{\mathrm{i}} n_{\mathrm{e}} + 
n_{\mathrm{i}} m_{\mathrm{e}})} \biggl(\frac{m_{\mathrm{i}}}{m_{\mathrm{e}}}\biggr) + 
\frac{\Gamma_{\mathrm{i}\gamma}}{e (m_{\mathrm{i}} n_{\mathrm{e}} + 
n_{\mathrm{i}} m_{\mathrm{e}})} \biggl(\frac{m_{\mathrm{e}}}{m_{\mathrm{i}}}\biggr)\biggr] \vec{J}\biggr\},
\label{COLL14}
\end{eqnarray}
where the plasma frequencies and the baryonic velocity have been introduced:
\begin{equation}
\omega_{\mathrm{p\,e,\,i}} = \sqrt{\frac{4 \pi n_{\mathrm{e,\,i}} e^2}{m_{\mathrm{e,\,i}} a}}, \qquad 
\vec{v}_{\mathrm{b}} = 
\frac{m_{\mathrm{e}} \vec{v}_{\mathrm{e}} + m_{\mathrm{p}} \vec{v}_{\mathrm{i}}}{m_{\mathrm{e}} + m_{\mathrm{p}}}.
\label{COLL14a}
\end{equation}
The evolution of $\vec{v}_{\mathrm{b}}$ is coupled to the velocity of the photons and it is obtained by summing up (instead of subtracting) Eq. (\ref{COLL5}) (multiplied by $m_{\mathrm{i}}$) and Eq. (\ref{COLL6}) (multiplied by $m_{\mathrm{e}}$):
\begin{eqnarray}
&& \vec{v}_{\mathrm{b}}^{\,\prime} + {\mathcal H}  \vec{v}_{\mathrm{b}} = \frac{\vec{J}\times 
\vec{B}}{a^4 \tilde{\rho}_{\mathrm{b}}(1 + m_{\mathrm{e}}/m_{\mathrm{i}})} - \vec{\nabla}\phi + \frac{4}{3}
 \frac{\tilde{\rho}_{\gamma}}{\tilde{\rho}_{\mathrm{b}}} a \Gamma_{\gamma\, \mathrm{e}} (\vec{v}_{\gamma} - \vec{v}_{\mathrm{b}}),
\label{COLL12}\\
&& \vec{v_{\gamma}}\,' = -\frac{1}{4} \vec{\nabla} \delta_{\gamma} - \vec{\nabla}\phi + a \Gamma_{\gamma\,\mathrm{e}} ( \vec{v}_{\mathrm{b}} - \vec{v}_{\gamma}).
\label{COLL13}
\end{eqnarray}
Eq. (\ref{COLL14}) can be expanded in power  in powers of $(m_{\mathrm{e}}/m_{\mathrm{i}})$. Recall that $\rho_{\mathrm{b}} = m_{\mathrm{e}} \tilde{n}_{\mathrm{e}} + m_{\mathrm{i}} \tilde{n}_{\mathrm{i}}$ and that, by global neutrality,  
 $n_{\mathrm{i}} = n_{\mathrm{e}} = n_{0}$
where $n_{0} = \eta_{\mathrm{b}} n_{\gamma}$ where $\eta_{\mathrm{b}}$ is the ratio between the baryonic 
concentration and the photon concentration already introduced after Eq. (\ref{con2}). The result of this double expansion implies, from Eq. (\ref{COLL14}),
\begin{eqnarray}
\frac{\partial \vec{J}}{\partial \tau} + \biggl({\mathcal H} + a\Gamma_{\mathrm{ie}} + 
\frac{4 \rho_{\gamma}\Gamma_{\mathrm{e}\gamma}}{ 3 n_{0} \,m_{\mathrm{e}}}\biggr) \vec{J} = \frac{\omega_{\mathrm{pe}}^2 }{4\pi} \biggl(\vec{E} + \vec{v}_{\mathrm{b}} \times \vec{B} + \frac{\vec{\nabla} p_{\mathrm{e}}}{e\,n_{0}}- \frac{\vec{J}\times 
\vec{B}}{e n_{0}}\biggr) + \frac{4 e \rho_{\gamma} \Gamma_{\mathrm{e}\gamma}}{3 m_{\mathrm{e}}} 
(\vec{v}_{\mathrm{b}} - \vec{v}_{\gamma}).
\label{COLL15}
\end{eqnarray}
The terms $\vec{J}'$ and ${\mathcal H} \vec{J}$ are 
comparable in magnitude and are both smaller than $\Gamma_{\mathrm{ie}}$ and 
$\Gamma_{\mathrm{e}\gamma}$, i.e. 
${\mathcal H} \vec{J} \simeq \vec{J}' < (4/3) (\rho_{\gamma}/m_{\mathrm{e}}) \Gamma_{\mathrm{e}\gamma} <a\Gamma_{\mathrm{ie}}$. While it is important to solve the evolution of $\vec{J}$ during all the pre-decoupling regime,  
the previous chain of inequalities implies that, asymptotically, the form of Eq. (\ref{COLL15}) is dominated 
by the term containing $\Gamma_{\mathrm{ie}}$. At the right-hand side the term 
containing $(\vec{v}_{\mathrm{b}} - \vec{v}_{\gamma})$ can be estimated by subtracting Eqs. (\ref{COLL12}) and (\ref{COLL13}). The difference $(\vec{v}_{\mathrm{b}} - \vec{v}_{\gamma})$ is driven exponentially to zero 
at a rate controlled by $a \Gamma_{\gamma\mathrm{e}} (1 + R_{\mathrm{b}}^{-1})$ where $R_{\mathrm{b}} = (3/4) \tilde{\rho_{\mathrm{b}}}/\tilde{\rho}_{\gamma}$. The asymptotic form of the Ohm's law 
can be written, for large conformal times as
\begin{equation}
\vec{J} = \frac{\omega_{\mathrm{pe}}^2 }{4\pi[a \Gamma_{\mathrm{ie}} + (4/3)(\rho_{\gamma}/m_{\mathrm{e}}) \Gamma_{\mathrm{e}\gamma}]} \biggl(\vec{E} + \vec{v}_{\mathrm{b}} \times \vec{B} + \frac{\vec{\nabla} p_{\mathrm{e}}}{e\,n_{0}}- \frac{\vec{J}\times \vec{B}}{n_{0} e}\biggr).
\label{APP4}
\end{equation}
The term containing the gradient of the electron pressure is the curved-space counterpart of the  thermoelectric term \cite{spitzer} while the term proportional to the vector product of the current and of the magnetic field is the curved-space counterpart of the Hall term.  The displacement current can be neglected in comparison 
with the Ohmic current (i.e.  $4\pi \vec{J} \gg \vec{E}\,'$) provided the left hand side of  Eq. (\ref{COLL15})  is subleading 
in comparison with the induced electric field (i.e. $\vec{J}' \ll \omega_{\mathrm{p\,e,i}}^2 \vec{E}$). 
The latter requirement demands, after derivation with respect to the conformal time $\tau$, the fulfillment of the condition 
$\vec{J}\,'' \ll \omega_{\mathrm{pe}}^2 \vec{J}$: the one-fluid 
description correctly captures the dynamics in the low-frequency branch of the spectrum of plasma excitations, i.e. 
$\omega \ll \omega_{\mathrm{pe}}$. 
If the thermoelectric and Hall terms are neglected, then the electromagnetic fields obey the following pair of equations, i.e. 
\begin{equation}
\frac{\partial \vec{B}}{\partial \tau} = \vec{\nabla} \times (\vec{v}_{\mathrm{b}} \times \vec{B})+ \frac{1}{4\pi \sigma} \nabla^2 \vec{B},\qquad  \frac{\partial \vec{E}}{\partial \tau} = -\frac{\partial}{\partial\tau} (\vec{v}_{\mathrm{b}} \times \vec{B})
+ \frac{1}{4\pi \sigma} \nabla^2 \vec{E},
\label{APP8}
\end{equation}
where $\sigma= \omega_{\mathrm{pe}}^2/(4\pi \Gamma_{\mathrm{ei}})$ is given by Eq. (\ref{COLL4});
Eq. (\ref{APP8}) implies that wavenumbers $k^2 > k_{\sigma}^2 =\sigma {\mathcal H}$ are dissipated because of the finite value of the conductivity.  The explicit value of the diffusivity scale $k_{\sigma}$ is given by
\begin{equation}
\frac{k_{\sigma}}{T} = 2.177 \times 10^{-18} \biggl(\frac{g_{\rho}}{10.75}\biggr)^{1/4} 
\biggl(\frac{h_{0}^2 \Omega_{\mathrm{M}0}}{0.1326}\biggr)^{3/4} \alpha^{-3/4}, \qquad \alpha = \frac{a}{a_{\mathrm{eq}}}.
\label{APP9}
\end{equation}
In comoving temperature units, the Hubble wavenumber is ${\mathcal H}/T = 1.047 \times  10^{-31} \sqrt{g_{\rho}/10.75}$ (where $g_{\rho}$ is the effective 
number of relativistic degrees of freedom); as expected not only ${\mathcal H} \ll k_{\sigma}$ but also 
${\mathcal H} \ll k_{\mathrm{Debye}}  \simeq 5.26 \times 10^{-6}\, T$ where $k_{\mathrm{Debye}}$ is 
the wavenumber corresponding to  $\lambda_{\mathrm{Debye}}=
\sqrt{T/(8 \pi n_{0} e^2)}$ i.e. the screening length of the Coulomb potential between 
two charges in the plasma. The diffusivity scale, on the contrary, sets a (lower) limit in the 
coherence scale of the Ohmic fields. The dominance of the drift term (i.e. $|\vec{v}_{\mathrm{b}} \times \vec{B}|$) over the thermoelectric and Hall terms demands, from Eq. (\ref{APP4}), the fulfillment of the following pair of relations:
\begin{equation}
\biggl(\frac{B}{\mathrm{nG}}\biggr) > 10^{8.27} \biggl(\frac{k}{T}\biggr) \biggl(\frac{h_{0}^2 \Omega_{\mathrm{M}0}}{0.1326}\biggr)^{-1/2} \, \sqrt{\alpha},\qquad  \biggl(\frac{B}{\mathrm{nG}}\biggr) < 10^{-11.19} \biggl(\frac{T}{k}\biggr)  
\biggl(\frac{h_{0}^2 \Omega_{\mathrm{M}0}}{0.1326}\biggr)^{1/2} \, \frac{1}{\sqrt{\alpha}}.
\label{APP11}
\end{equation}
\begin{figure}[!ht]
\centering
\includegraphics[height=6cm]{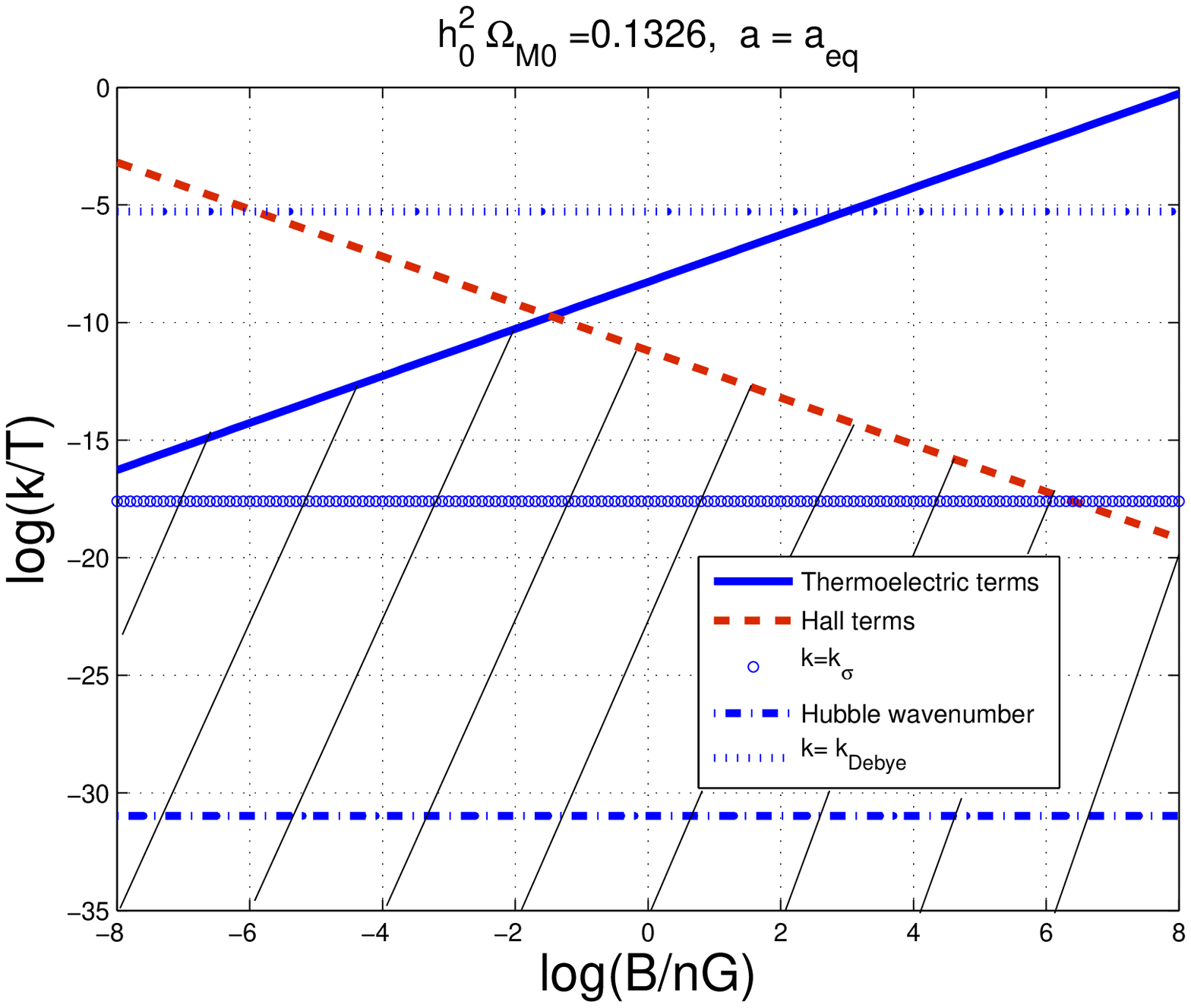}
\includegraphics[height=6.1cm]{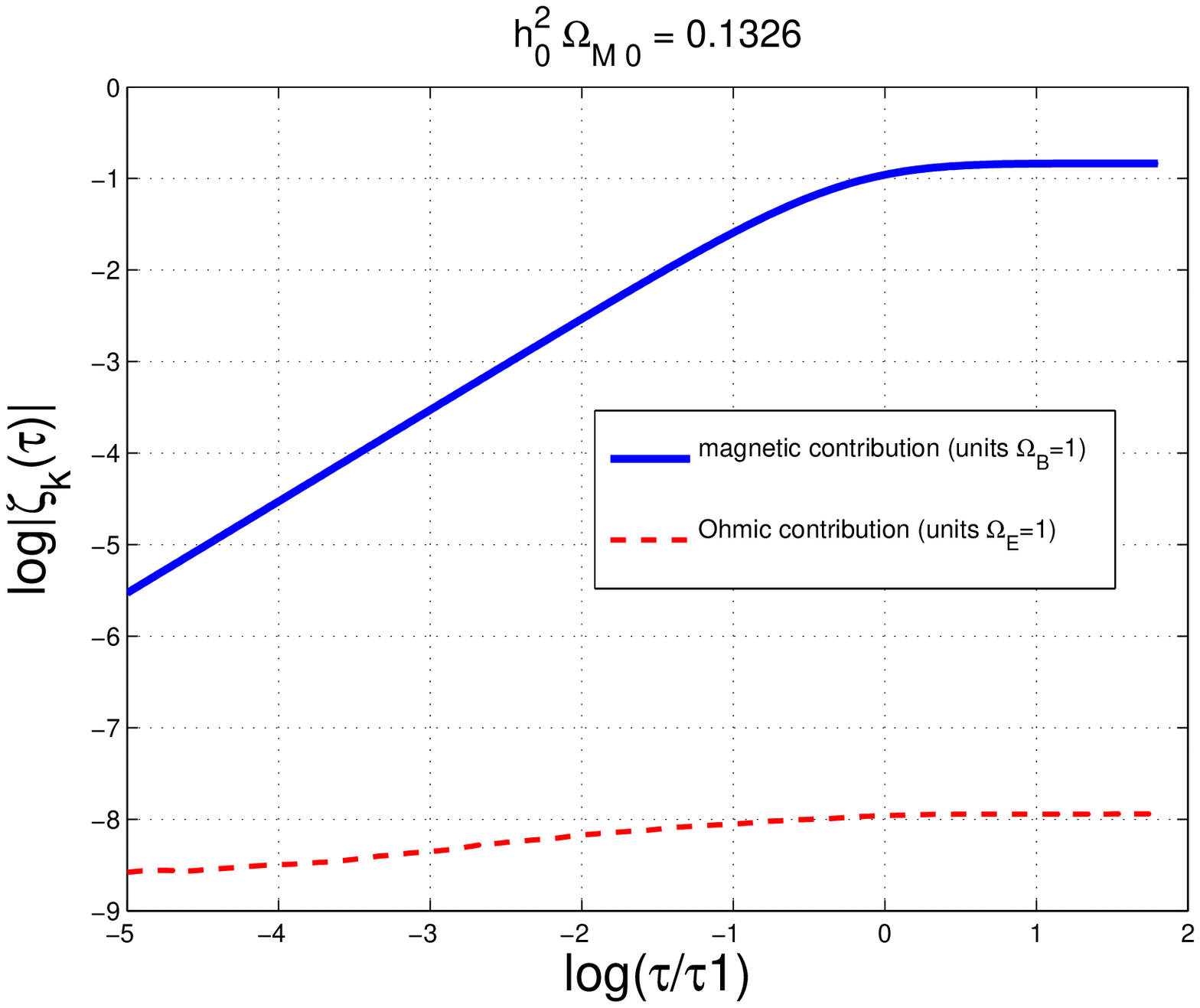}
\caption[a]{Relative contribution of Hall, thermoelectric and drift terms in the Ohmic current (plot at the left). Contribution 
of Ohmic currents to the evolution of curvature perturbations (plot at the right). In the left plot (as well as in Fig. \ref{FIG2}) 
$\tau_{1}$ sets the equality time scale (more precisely $\tau_{\mathrm{eq}} = (\sqrt{2} -1)\tau_{1}$).}
\label{FIG1}      
\end{figure}
In Fig. \ref{FIG1} (plot at the left) the contribution of the Hall and thermoelectric terms 
to the Ohmic current is illustrated.  For comparison the diffusivity, Debye and Hubble scales
are also reported. The shaded area denotes the region where the conditions of Eq. (\ref{APP11}) 
are approximately fulfilled, i.e. for typical amplitudes of the comoving magnetic field 
in the range $10^{-5} \mathrm{nG} < B< 10^{5} \mathrm{nG}$. Because 
of the value of the charge concentration, the Hall contribution dominates over  the  thermoelectric term of the electrons
\begin{equation}
\frac{|\vec{J}\times \vec{B}|}{|\vec{\nabla} p_{\mathrm{e}}|} \simeq \frac{B^2}{4 \pi n_{0} T}  \simeq 
8.19 \times 10^{2} \biggl(\frac{B}{\mathrm{nG}}\biggr)^2 
\biggl(\frac{h_{0}^2 \Omega_{\mathrm{b}0}}{0.02273}\biggr).
\label{OM2}
\end{equation}
\begin{figure}[!ht]
\centering
\includegraphics[height=6cm]{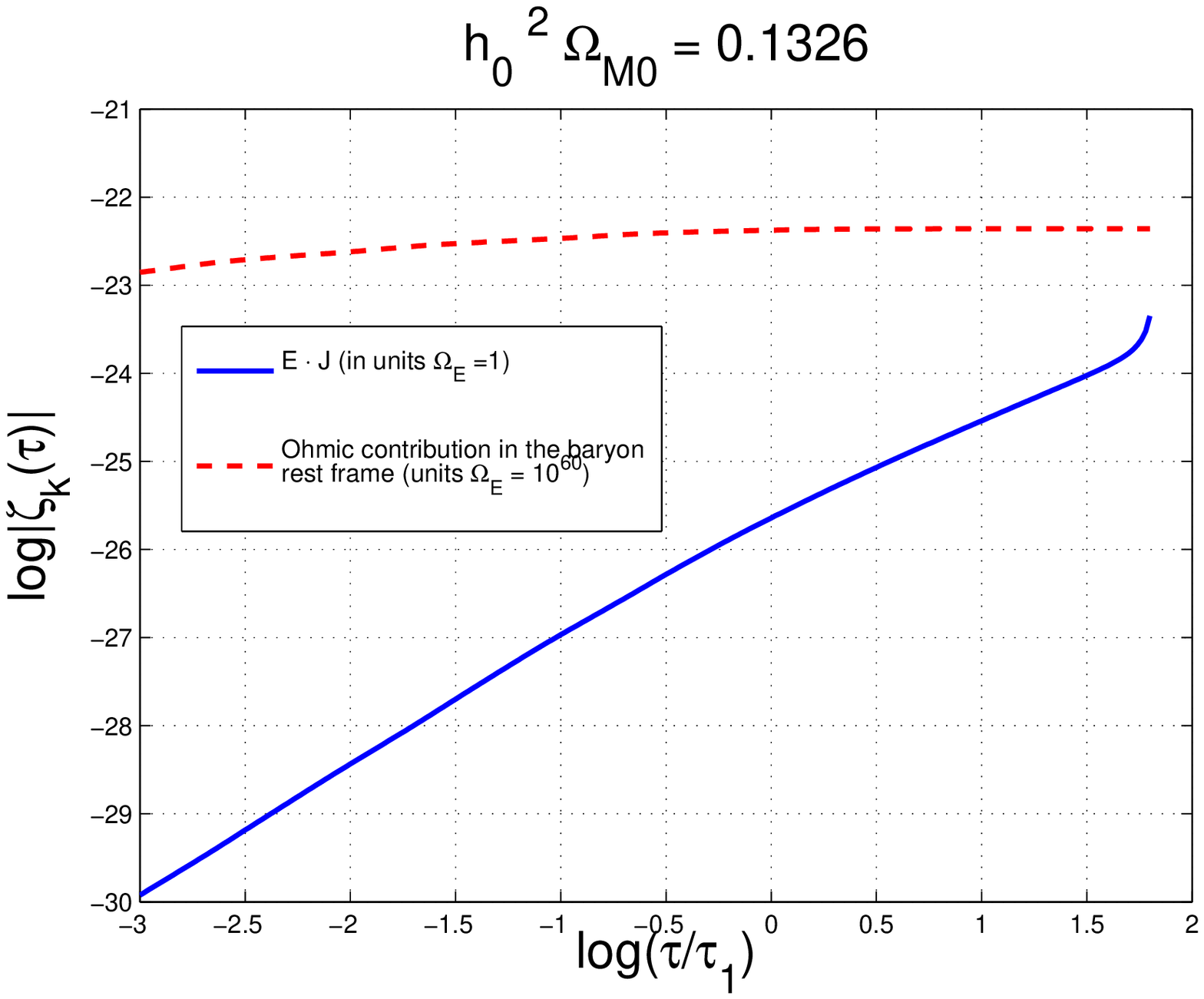}
\includegraphics[height=6cm]{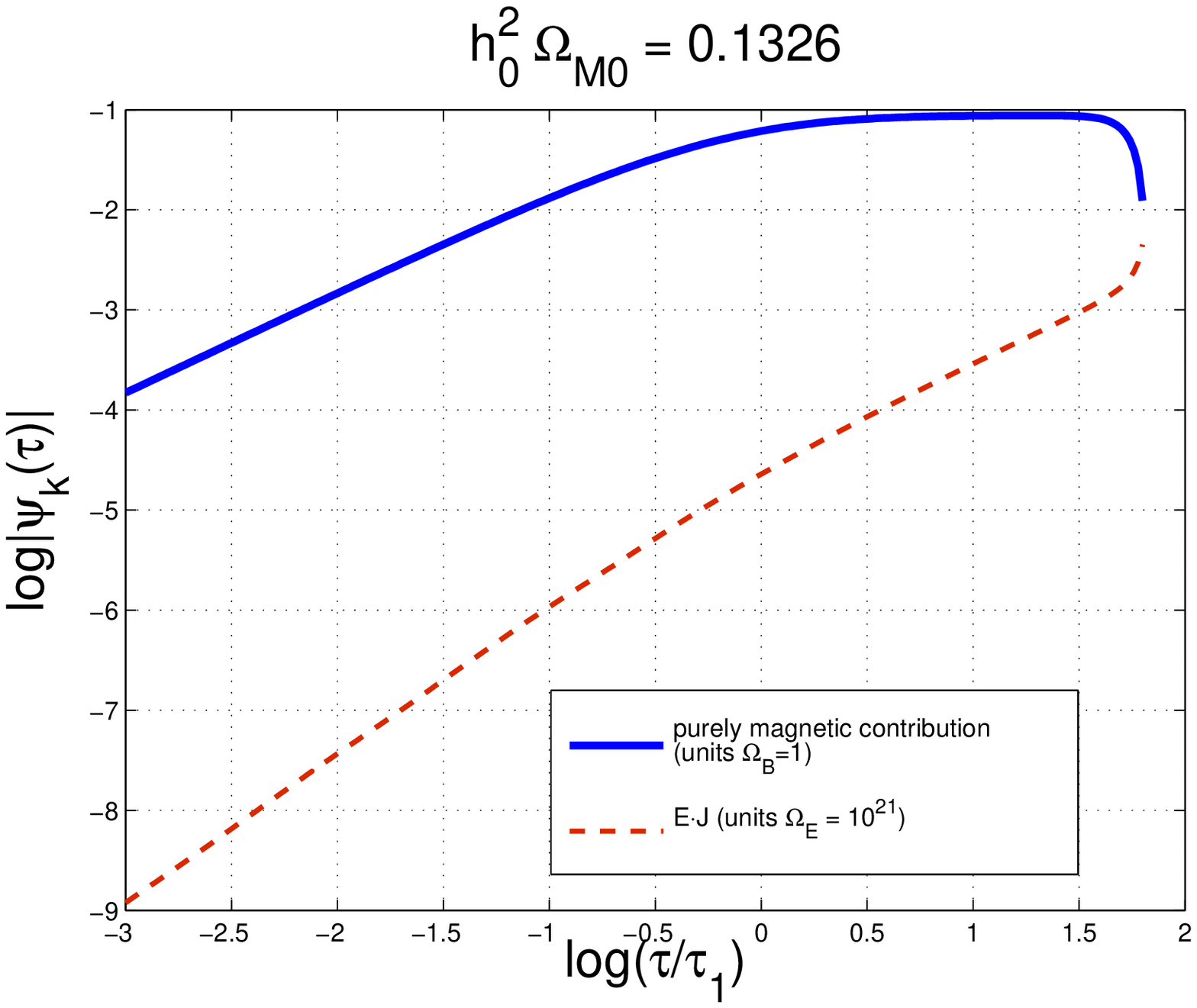}
\caption[a]{The contribution of the Ohmic fields to the curvature perturbations (plot at the left) and to the metric 
fluctuations (plot at the right).}
\label{FIG2}      
\end{figure}
When the plasma contains a magnetic field whose Fourier modes are stochastically distributed with power spectrum $P_{\mathrm{B}}(k)$, the asymptotic form of the Ohm law (\ref{APP4}), within the shaded region of the parameter space of Fig. \ref{FIG1}, 
induces an effective electric field 
\begin{eqnarray}
&& \langle E_{i} (\vec{k},\tau) E_{j}(\vec{p},\tau) \rangle = \frac{2\pi^2}{k^3} P_{ij}(k) P_{\mathrm{E}}(k,\tau)  \delta^{(3)}(\vec{k} + \vec{p}), \qquad P_{\mathrm{E}}(k,\tau) = 
\biggl[ g_{1}^2  \alpha(\tau) + \frac{g_{2}^2}{\alpha(\tau)}\biggr] P_{\mathrm{B}}(k)
\nonumber\\
&& g_{1} = 4.29 \times 10^{-10} \biggl(\frac{k}{T}\biggr) \biggl(\frac{h_{0}^2 \Omega_{\mathrm{M}0}}{0.1326} \biggr)^{-1/2},\qquad 
 g_{2} = 4.89 \times 10^{-5} \biggl(\frac{h_{0}^2 \Omega_{\mathrm{M}0}}{0.1326} \biggr)^{1/2}.
\label{zeta4c} 
\end{eqnarray}
where $P_{ij}(k) = (\delta_{ij} - k_{i}k_{j}/|\vec{k}|^2)$ is the transverse projector. The stochastic electromagnetic 
fields as well as the induced Ohmic currents, being inhomogeneous, affect the curvature perturbations \footnote{We shall denote by $\zeta$ the density contrast  on uniform curvature hypersurfaces (see, e. g. \cite{UN})
while ${\mathcal R}$ is the curvature perturbation on comoving orthogonal hypersurfaces. The two quantities are connected 
by the Hamiltonian constraint, i.e. Eq. (\ref{zeta2}).}
whose evolution, on the absence of non-adiabatic pressure fluctuations, 
is given by \cite{max2,max3}:
\begin{eqnarray}
\zeta'  = \frac{\vec{E} \cdot \vec{J}}{3 a^4 (\tilde{p}_{\mathrm{t}} + \tilde{\rho}_{\mathrm{t}})} + \frac{{\mathcal H}\delta_{\mathrm{s}} \rho_{\mathrm{B}} (3 c_{\mathrm{st}}^2 -1)}{ 3 (\tilde{\rho}_{\mathrm{t}} + \tilde{p}_{\mathrm{t}})} + \frac{{\mathcal H} \delta_{\mathrm{s}} \rho_{\mathrm{E}}}{ 3 (\tilde{\rho}_{\mathrm{t}} + \tilde{p}_{\mathrm{t}})} \frac{[3 c_{\mathrm{st}}^2 g_{1}^2 \alpha^2 + g_{2}^2 ( 3
c_{\mathrm{st}}^2 -2)]}{g_{1}^2 \alpha^2 +g_{2}^2}
 - \frac{(\vec{\nabla} \cdot \vec{v}_{\mathrm{t}})}{3},
\label{zeta1}
\end{eqnarray}
where $\delta_{\mathrm{s}} \rho_{\mathrm{B}} = B^2/(8\pi a^4)$ and  $\delta_{\mathrm{s}} \rho_{\mathrm{E}} = E^2/(8\pi a^4)$; moreover  
\begin{equation}
\zeta = {\mathcal R} + \frac{\nabla^2 \psi}{12 \pi G a^2 ( \tilde{p}_{\mathrm{t}} + \tilde{\rho}_{\mathrm{t}})}, \qquad 
{\mathcal R} = - \psi - \frac{{\mathcal H} ({\mathcal H} \phi + \psi')}{{\mathcal H}^2 - {\mathcal H}'}.
\label{zeta2}
\end{equation}
Note that $(\tilde{p}_{\mathrm{t}} + \tilde{\rho}_{\mathrm{t}}) \vec{v}_{\mathrm{t}} = \sum_{a} (\tilde{p}_{\mathrm{a}} + \tilde{\rho}_{\mathrm{a}}) \vec{v}_{\mathrm{a}}$
is the total velocity field of the plasma including the contribution of cold dark matter particles, neutrinos, 
electrons, ions and photons. 
When the Universe contains matter, radiation and dark energy the total barotropic index $w_{\mathrm{t}}$ and the total 
 sound speed $c_{\mathrm{st}}^2$ can be written, respectively, as 
\begin{eqnarray}
w_{\mathrm{t}} =  \frac{\tilde{p}_{\mathrm{t}}}{\tilde{\rho}_{\mathrm{t}}} =\frac{\alpha_{\Lambda}^{3} - 3 \alpha^4}{3 ( \alpha_{\Lambda}^3 + \alpha^4 + \alpha \alpha_{\Lambda}^3)},
\qquad c_{\mathrm{st}}^2 = \frac{\tilde{p}_{\mathrm{t}}'}{\tilde{\rho}_{\mathrm{t}}'} = w_{\mathrm{t}}  - \frac{\alpha}{3(1 + w_{\mathrm{t}})} \frac{\partial w_{\mathrm{t}}}{\partial \alpha},
\label{zeta8}
\end{eqnarray}
where, as already stressed, $\alpha = a/a_{\mathrm{eq}}$; furthermore $\tilde{\rho}_{\mathrm{t}}
= \tilde{\rho}_{\mathrm{R}} + \tilde{\rho}_{\mathrm{M}} + \tilde{\rho}_{\Lambda}$ and 
\begin{equation}
\alpha_{\Lambda} = \frac{a_{\Lambda}}{a_{\mathrm{eq}}} = 2246.81 \biggl(\frac{h_{0}^2 \Omega_{\mathrm{M}0}}{0.1326}
\biggr)^{4/3} \biggl(\frac{h_{0}^2 \Omega_{\Lambda}}{0.3835}\biggr)^{-1/3},
\qquad \alpha_{0} = 3195.18\, \biggl(\frac{h_{0}^2 \Omega_{\mathrm{M}0}}{0.1326}\biggr).
\label{zeta9}
\end{equation}
The evolution of $\zeta$ as well as the evolution of $\psi$
can be integrated and the results are reported in Fig. \ref{FIG1} (plot at the left) and in Fig. \ref{FIG2}.
Both in Fig. \ref{FIG1} and \ref{FIG2} $\Omega_{\mathrm{B}} = \delta_{\mathrm{s}} \rho_{\mathrm{B}}/(8\pi \tilde{\rho}_{\gamma})$ and 
$\Omega_{\mathrm{E}} = \delta_{\mathrm{s}} \rho_{\mathrm{E}}/(8\pi \tilde{\rho}_{\gamma})$. In Fig. \ref{FIG1} (plot at the left) 
the pure magnetic contribution is compared to the total Ohmic contribution in units $\Omega_{\mathrm{E}} =1$.  In 
 Fig. \ref{FIG2} the contributions of $\vec{E}\cdot\vec{J}$ is compared to the other terms arising in the evolution of $\zeta$ and $\psi$. In the baryon rest 
 frame the Ohmic contribution is suppressed as $(k^2/\sigma^2) \sim {\mathcal O}(10^{-60})$ for length-scales larger 
 than the Hubble radius (notice that, indeed, in the left plot of Fig. \ref{FIG2} $\Omega_{\mathrm{E}}$ has been rescaled by a factor 
 $10^{60}$   to make the two contribution visually comparable on a linear scale). The latter result is compared 
 with the suppression experience by $\vec{E}\cdot\vec{J}$ which is of the order of $(k^2/k_{\sigma}^2) \sim {\mathcal O}(10^{-21})$.
In Fig. \ref{FIG2} (plot at the right) the contribution of $\vec{E}\cdot\vec{J}$ (rescaled by a factor $10^{21}$) is compared 
with the magnetic contribution as it arises in Eq. (\ref{zeta1}). The 
Ohmic contribution is dominated by the drift term which vanishes in the baryon rest frame and which is subleading over 
typical length-scales larger than the Hubble radius.  An interesting byproduct of this study is the derivation of a  consistent evolution equation 
for the Ohmic current. The latter result improves on the usual approximations posited in  
 the Boltzmann integrators  accounting for the effects of large-scale magnetic fields \cite{max2,max3} on CMB observables.  
 
NQL wishes to thank the CERN physics department and, in  particular, 
 Prof. L. Alvarez-Gaum\'e for kind hospitality and financial support. 

\end{document}